\documentclass[aps,prl,showkeys,showpacs,preprint,groupedaddress]{revtex4}
\usepackage{graphicx}

\begin{document}

\title{Photons uncertainty removes Einstein-Podolsky-Rosen paradox}

\author{Daniele Tommasini}

\affiliation{Departamento de F\'\i sica Aplicada, \'Area de
F\'\i sica Te\'orica, Universidad de Vigo, 32004 Ourense, Spain}

\email[]{daniele@uvigo.es}

\date{\today}

\begin{abstract}
Einstein, Podolsky and Rosen (EPR) argued that the
quantum-mechanical probabilistic description of physical reality
had to be incomplete, in order to avoid an instantaneous action
between distant measurements. This suggested the need for
additional ``hidden variables", allowing for the recovery of
determinism and locality, but such a solution has been disproved
experimentally. Here, I present an opposite solution, based on the
greater indeterminism of the modern quantum theory of Particle
Physics, predicting that the number of photons is always
uncertain. No violation of locality is allowed for the physical
reality, and the theory can fulfill the EPR criterion of
completeness.
\end{abstract}

\pacs{03.65.-w; 03.65.Ta; 11.10.-z; 11.15.-q; 12.20-m}

\keywords{Quantum Field Theory; Standard Model; Quantum
Electrodynamics; Soft Photons; Quantum Measurement; EPR paradox}


\maketitle

{\bf Introduction.} In the abstract of their original 1935 paper
(probably the most cited paper in the history of physics),
Einstein, Podolsky and Rosen (EPR) summarize their argument as
follows: ``In a complete theory there is an element corresponding
to each element of reality. A sufficient condition for the reality
of a physical quantity is the possibility of predicting it with
certainty, without disturbing the system. In quantum mechanics in
the case of two physical quantities described by non-commuting
operators, the knowledge of one precludes the knowledge of the
other. Then either (1) the description of reality given by the
wave function in quantum mechanics is not complete or (2) these
two quantities cannot have simultaneous reality. Consideration of
the problem of making predictions concerning a system on the basis
of measurements made on another system that had previously
interacted with it leads to the result that if (1) is false then
(2) is also false. One is thus led to conclude that the
description of reality as given by a wave function is not
complete" \cite{EPR}.

As Lalo\"e noticed in a recent review, for Einstein and
collaborators ``the basic motivation was not to invent paradoxes;
it was to build a strong logical reasoning which, starting from
well-defined assumptions (roughly speaking: locality and some form
of realism), would lead ineluctably to a clear conclusion (quantum
mechanics is incomplete, and even: physics is deterministic)"
\cite{Laloe}. In fact, the EPR argument has been one of the main
motivations for seeking a ``complete" deterministic theory
underlying quantum mechanics. In their (thought) ``EPR
experiment", the measurement of a physical quantity on a system A
did influence instantaneously, and in a {\it perfectly
deterministic} way, the result of a corresponding measurement
performed on another spatially separated system B that had been
interacting with A in the past. This fact, which was since called
the EPR paradox, was considered to be a hint for a fully
deterministic theory underlying the probabilistic quantum theory.

However, we will see that the modern Quantum Field Theory (QFT)
description of Particle Physics, known as the Standard Model (SM)
\cite{WeinbookI,WeinbookII}, predicts a fundamental uncertainty
about the number of photons that can be produced in any process
involving an interaction. As a consequence, two distant
measurements cannot influence each other in a certain,
deterministic way, as required by the EPR argument.

{\bf The EPR thought-experiment and the EPR paradox.} For our
purposes, it will be sufficient to consider a class of ``EPR
experiments" defined as follows: two particles A and B
\footnote{It is easy to generalize the present discussion to the
case of three or more particles.} are emitted by a source; far
apart, some conserved observable, such as a component of angular
momentum (spin, helicity or polarization), is measured on particle
A. According to the usual quantum-mechanical treatment, the
measurement carried out on A reduces its state into an eigenstate
of the measured observable, whose conservation immediately forces
the second particle (B) to ``collapse" into a corresponding
eigenstate of this observable as well. For instance, let A and B
be two spin 1/2 particles, produced with zero total angular
momentum in a singlet spin state, described by the ``entangled"
spin vector
\begin{equation}
\vert\psi\rangle=\frac{1}{\sqrt{2}}\left(\vert+\rangle_{A}\vert-\rangle_{B}
- \vert-\rangle_{A}\vert+\rangle_{B}\right), \label{entanglement}
\end{equation}
where $\vert\pm\rangle_A$ are the usual eigenstates of the spin
component $S_z(A)$ of particle A with eigenvalues $\pm\hbar/2$
respectively (this example is usually called the ``EPR-Bohm"
experiment \cite{Bohm,Laloe}).

Assuming the state of Eq.˜ (\ref{entanglement}), it is easy to see
that the measurement of the spin component $S_z$ performed
independently on any of the two particles can give both values
$\pm\hbar/2$, each with probability 1/2. On the other hand, if in
a given single event $S_z(A)$ is measured on A and found equal to,
say, $+\hbar/2$, it will then be possible to {\it predict with
certainty} the result of the measurement of $S_z(B)$ on the
distant particle B, that will give $-\hbar/2$. The quantity $S_z$
observed on A would then instantaneously acquire ``an element of
physical reality" also on B, according to the original definition:
{\it ``If, without in any way disturbing a system we can predict
with certainty (i.e.˜ with probability equal to unity) the value
of a physical quantity, then there is an element of physical
reality corresponding to this physical quantity"} \cite{EPR} (the
italics and the parenthesis also belong to the original paper).
Moreover, the element of physical reality on B depends on the
actual measurement that is done on A: for instance, if instead of
measuring the component $S_z(A)$ of the angular momentum we
decided to measure an observable incompatible with it, such as the
component $S_x(A)$, then the state of the distant particle B after
such a distant measurement would become an eigenstate of $S_x(B)$,
rather than one of $S_z(B)$. Therefore, assuming that the two
systems are no longer interacting, EPR deduced that two
quantum-mechanically incompatible quantities ($S_z(B)$ and
$S_x(B)$ in the example above) could be given a simultaneous
reality. They then concluded that ``the wave function does not
provide a complete description of the physical reality";
otherwise, the definite values of $S_z(B)$ and $S_x(B)$ would have
to ``enter into the complete description, according to the
condition of completeness" that they had defined as follows: {\it
``every element of the physical reality must have a counterpart in
the physical theory"} \cite{EPR}.

The EPR argument (later called paradox, see e.g.˜ Ref.˜
\cite{Einstein}) is so rigorous, that Lalo\"e reformulated it in
the form of a theorem: {\it ``If the predictions of quantum
mechanics are correct (even for systems made of remote correlated
particles) and if physical reality can be described in a local (or
separable) way, then quantum mechanics is necessarily incomplete:
some `elements of reality' exist in Nature that are ignored by
this theory"} \cite{Laloe}. The importance of this argument, based
on their objective definition of physical reality, is that it
holds almost independently of the interpretation of the theory,
with the exception of deterministic interpretations introducing
hidden variables \cite{Laloe}. It is then impossible to find an
``EPR-paradox-free" interpretation of the probabilistic quantum
mechanics based on Eq.˜ (\ref{entanglement}). It is the latter
equation, i.e.˜ the entangled state vector description, which
should be rejected, as EPR pointed out.

What EPR did not perhaps expect was that a way out was to be found
in the modern version of the Quantum Theory itself. In fact, as we
shall see, the SM description does not rely on the ``entangled"
state of A and B (whose spin part is Eq.˜ (\ref{entanglement}) in
the example that we have considered above), but it allows for the
presence of an undetermined number of additional photons.

{\bf The uncertainty about the number of photons.} The EPR
paradox, as described above, originates from the assumption of a
state with a definite number of particles (two in our example),
which is incorrect in Relativistic Quantum Mechanics. As we shall
see, {\it the modern QFT description of Particle Physics predicts
that it is impossible to prepare a state with a definite number of
particles as the result of a given physical process, since
additional real particles can always be created in the production
process itself.} Which additional species can appear depends on
the available energy. Since massless particles can have
arbitrarily low energy, the possible presence of real ``soft
photons" (i.e.˜ photons having a low enough energy) should always
be taken into account in the theoretical treatment.

\begin{figure}
\includegraphics{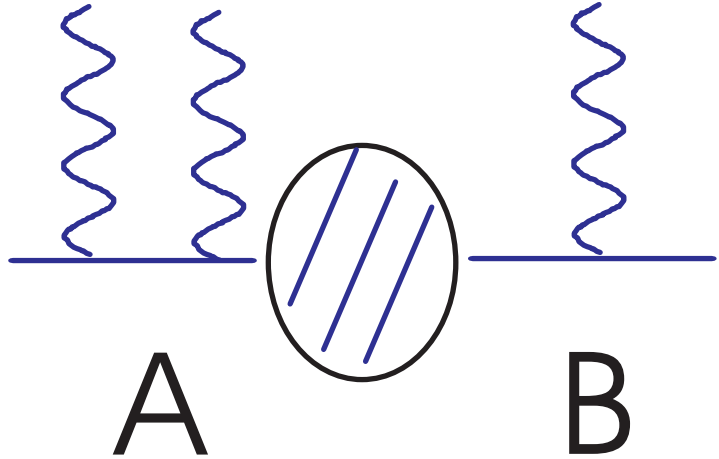}
\caption{\label{fig1} Feynman diagram describing the production of
an EPR pair of charged spin 1/2 particles, A and B, in coincidence
with three additional photons. The dashed blob represents the part
depending on the particular basic process and the initial
particles that are considered.}
\end{figure}

\begin{figure}
\includegraphics{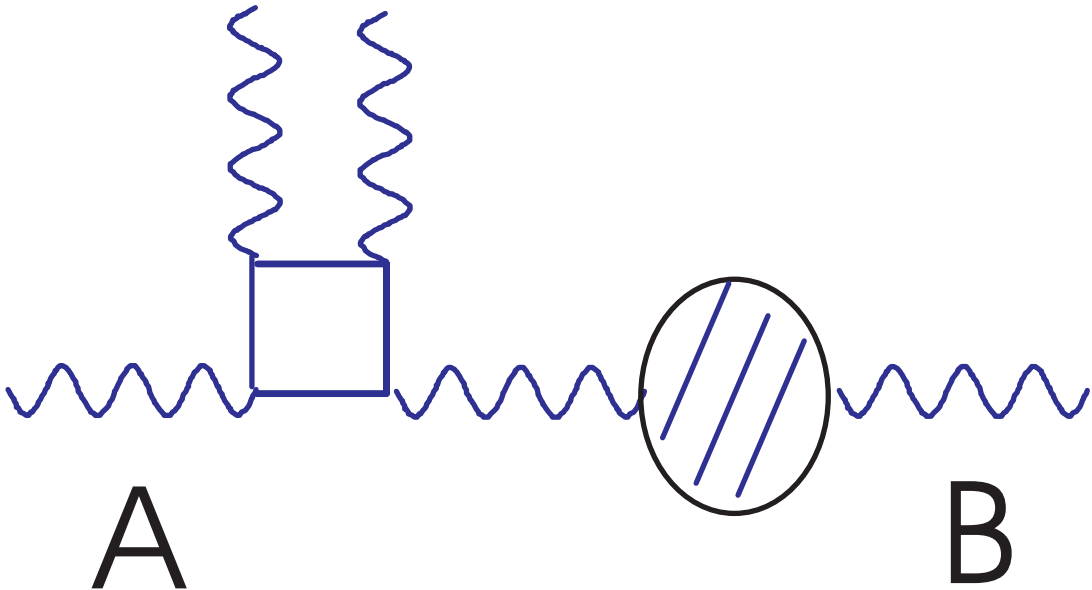}
\caption{\label{fig3} Feynman diagram describing the production of
an EPR pair of photons, A and B, in coincidence with two
additional photons. The dashed blob represents the part depending
on the particular basic process and the initial particles that are
considered.}
\end{figure}

Here, I will prove this statement using QFT perturbation theory
(i.e.˜ Feynman diagrams). To be concrete, I will first discuss two
kinds of ideal EPR experiments: i) those involving two charged
spin 1/2 particles; and ii) those involving two photons. In both
cases, I will give explicit examples predicting the creation of an
arbitrary number of additional photons.

i) In Fig.˜ \ref{fig1}, I have drawn a tree-level diagram where
the ``blob" represents the particular elementary process that
produces particles A and B.
Even without specifying that part of the diagram (involving some
``initial" particles), we see that an arbitrary number of real
photons (three in the particular case of the figure) can be
attached to each of the external fermion legs (see also Chap. 13
of Ref.˜ \cite{WeinbookI}).

ii) Since no three photon vertex exists at the tree level, in the
``two-photon" EPR experiment we have to look for one loop effects.
In Fig.˜ \ref{fig3}, I show a ``box" diagram for the production of
two additional real photons \footnote{This box diagram has been
studied in different contexts, e.g.˜ in the theory of two photon
scattering.}.
The virtual particle in the loop can be any charged fermion
(electron, muon, tau, quarks).

Note that additional photons can also be emitted by the other legs
of the diagrams, those corresponding to the particles that are
included in the ``blob", or even by loop diagrams involving
charged particles that can contribute to the ``vertex part" of the
``blob". All the relevant SM diagrams can be drawn, depending on
the particular production process that is considered, although
Feynman perturbation theory breaks down when the additional
photons are attached to particles belonging to a bound system such
as an atom \cite{WeinbookI}. Fortunately, for the purposes of the
present paper I do not need an exact computation of all the
possible contributions.

The rates for the production of a given number of additional
photons should be compared to the rate for the process in which
only particle A and B are emitted, corresponding to the bare
diagram without any additional photons attached (that would imply
the same correlations as the old quantum-mechanical approach)
\footnote{The ``Infrared Divergencies" related to the integration
over arbitrarily small photons momenta can be handled as shown in
Chapter 13 of Ref.˜ \cite{WeinbookI}, and they eventually
cancel.}. In any case, such rates are suppressed by increasing
powers of the fine structure constant $\alpha\simeq 1/137$,
depending on the considered number of additional photons. Since
particle A is detected when a measure (for instance of some
angular momentum) is made on it, its energy $E_A$ can also be
measured. Therefore the upper limit for the total energy of the
additional photons is $\Lambda=E-E_A-m_Bc^2$ (where $E$ is the
total energy liberated in the basic production process). This
limit reduces the phase space available for the diagrams involving
an increasing number of additional photons. In the case of the
EPR-Bohm experiment, we will be interested in the diagrams
allowing for parallel (rather than antiparallel) spins of A and B.
Such diagrams are suppressed by powers of
$\left(\frac{\Lambda}{E}\right)^2$ (thus in the limit for
$\Lambda\to0$, the helicities of the two fermion will remain
opposite \cite{WeinbookI}), although this is not necessarily a
small factor in the EPR ideal experiment. In the case of a
two-photons EPR experiment, the probability of a diagram such as
that in Fig.˜ \ref{fig3} is suppressed by four powers of the fine
structure constant and by the electron propagators in the loop
(which are larger than the available energy, unless one considers
ideal EPR experiments with very energetic photons, at or above the
MeV range); the possible photons radiation from the charged
fermions that appear in the ``blob" part of the diagram could then
be more important (although it may not be computable by Feynman
perturbation theory).

To summarize, the precise suppression factor depends on the
particular case that is considered. For our purposes, it is
sufficient to note that there is a non-vanishing probability for
additional photons to be created, at least due to diagrams such as
those of Figs.˜ \ref{fig1} or \ref{fig3}, and that they contribute
to energy, momentum and angular momentum conservation.

The important fact is that this uncertainty principle can be
generalized: {\it an undetermined number of photons is created in
any experiment, in any step that involves an interaction,} and in
particular in the process that originates our EPR particles
\footnote{An undetermined number of photons can also be created
due to the interaction of any observed system with the particles
belonging to the measuring apparatus. Although such an effect will
not be used in the following discussion, it can be interesting for
the Theory of Measurement.}. In fact, all the known elementary
particles, including the neutral ones, such as the photon, the
neutrino (and even the possible Higgs boson), can radiate photons
when they appear as external legs of an interaction process. For
the charged particles and for the photon, this fact is shown in
Figs.˜ \ref{fig1} and \ref{fig3}. For a neutrino (or a Higgs
boson), it is easy to construct loop diagrams involving virtual W
bosons and charged fermions to which a photon line can be
attached. In the case of composite particles, such as the neutral
K or B mesons, the production process involves the constituent
quarks, charged particles to which external photon lines can be
attached just as in Fig.˜ \ref{fig1} \footnote{In all such cases,
additional photons can also be radiated by the lines of all the
possible charged particles (besides the EPR pair) that are
involved in the relevant ``blob" production process.}.

This generality is by no means accidental, but it corresponds to a
well-known characteristic of QFT: any process that does not
violate the fundamental symmetries is ``allowed" and has a
non-vanishing amplitude. Exceptions to such a ``rule" are so rare,
that they are thought to hide new symmetries. Here, it is
sufficient to note that no symmetry forbids the radiation of
additional photons in coincidence with a given interaction, since
photons do not carry any conserved ``internal" charge.

The uncertainty about photons radiation can also be related to a
symmetry principle. In fact, it is based on two points: the
existence of massless neutral particles, the photons; and of the
fermion-photon vertex, that allows for the radiation of photons by
any external line of the relevant Feynman diagrams (possibly
through loops as in Fig.˜ \ref{fig3}). But it is well known that
both the electromagnetic vertex and the masslessness of the photon
are the direct consequences of the local, unbroken
(electromagnetic) gauge symmetry.

{\bf The EPR paradox removed.} According to the previous
discussion, the state arising from the interaction is {\it never}
an eigenstate of the operator counting the number of photons: {\it
the number of photons cannot be determined (it never gets a
physical reality).} This implies that {\it it is never correct to
use a state with a fixed number of particles, such as that of Eq.˜
(\ref{entanglement}), as emerging from a given interaction.}

Now, in a given Feynman diagram, the conservation laws hold for
the set including particles A and B together with all the
additional photons that appear in that diagram. Therefore, {\it
after the measurement on A in any given single event, the energy,
momentum and angular momentum conservation laws do not hold for
the two particle (sub)system, A and B}. The detection of particle
A does not necessarily correspond to particle B appearing in the
opposite direction. Moreover, {\it the measurement on A does not
allow for a certain prediction of the value of the considered
conserved quantity} (be it energy, momentum or angular momentum)
{on B (B is not put in an eigenstate of the observable that has
been measured on A).} For instance, in the EPR-Bohm experiment,
$S_z$ {\it will not be given a ``physical reality" on B after it
is measured on the distant particle A.} According to our previous
discussion, this is sufficient to save the theory from the
original EPR paradox. Note that this result holds even for a small
probability of additional photons radiation.

Note that we can know that particle B appears in the given region,
and that it is in a given eigenstate of the considered observable,
only {\it after} detecting the particle B and measuring the
considered observable on it. Therefore, the physical reality is
given to the observables on B only after the measurement on its
own location is performed. {\it In this sense, QFT respects local
realism: the elements of physical reality of the theory can be
obtained only after local measurements.} According to this
definition, we can even say that QFT is locally realistic,
although such a definition is usually reserved to hidden variables
theories.

A general single event, where only particles A and B are detected,
can show apparent symmetry violations. In particular, any
violation of a discrete variable such as angular momentum is
important, since it is a multiple of $\hbar$. These considerations
suggest that a possible signature of the theoretical solution I am
proposing would be the observation of an apparent symmetry
violation event in an EPR experiment (actually due to the presence
of additional unobserved photons).

It is worth pointing out that there is no possibility of getting
rid of the uncertainty about the number of photons. Even if we
filled the whole space with detectors, we would never catch all
the possible photons involved in a single event, since they can
have arbitrarily low energy \cite{WeinbookI}. Only a definite
number of photons will be detected, while the state produced in
the basic process has no definite photons content. After the
measurement, the amplitude for the additional undetected photons
spreads over the whole space, eventually overlapping with A and B;
therefore there is no theoretical possibility to define two
determined spatially separated subsystems as required by the EPR
argument. Strictly speaking, in QFT it can be correct to say that
A and B themselves are spatially separated only {\it after}
measuring on both particles, since the measurement on A and global
momentum conservation are not sufficient to ensure the ``collapse"
of B as a particle in a given direction as well (it is even
possible that A and B are caught by the same detector!).

In other words, the modern QFT description is even less
deterministic than the old non-relativistic Quantum Mechanics. In
fact, the only predictions that it allows are on probabilities and
average values. This greater uncertainty protects the theory from
the EPR paradox. It seems that, to remove the paradox, one has to
choose between the most extreme possibilities: determinism (hidden
variables), or complete lack of determinism for the single event
(QFT, the dice of God) \footnote{However, the QFT field equations
are deterministic. This is an important point, whose possible
consequences for the problem of locality will be discussed
elsewhere.}. We also see that the SM fulfills the EPR criterion of
completeness that was cited above: for instance, in QFT the Fock
space state vector after the measurement of a conserved charge on
a massive particle is an eigenvector of that one-particle charge
operator. The element of physical reality that appears after the
measurement then has a counterpart in the theory, being the
corresponding eigenvalue (i.e.˜ the definite value) of this
one-particle observable \footnote{Possibly the only observables
that can actually get a local reality in QFT are the conserved
gauge charges. This would not prevent QFT to fulfill the EPR
criterion of reality, that only requires the existence of a
counterpart for all the physical reality of the theory.}.

Ultimately, the recovery of local realism can be attributed to the
local gauge symmetry that implies the uncertainty about the
photons as we have seen. I think that this result is not
surprising, since the local gauge symmetry implies local
interactions. Note also that the possibility of radiating
arbitrarily soft photons corresponds to the ``infrared" behavior
of the theory, i.e.˜ to its long distance properties, which are
precisely those that are expected to be relevant for the
discussion of distant measurements.

{\bf On Bell's variant of the EPR experiment.} The actual
so-called EPR experiments \cite{Aspect,Laloe}, that have been
inspired by the work of Bell \cite{Bell}, do not test the EPR
argument directly. First of all, the measurement on particle B is
performed and only the events where particles A and B appear in
coincidence at opposite directions are considered. The fact that
we need to detect B in order to define such events already
differentiates such experiments from the ideal EPR experiment.
Moreover, the results are given in terms of the correlations
between the polarizations of the two (or more) particles A and B,
which are statistical averages over the products of the observed
spin/polarization components for the different single events. For
instance, in Bell's version of the EPR-Bohm experiment, the
relevant correlations are the average values of the products of
the components $S_{\vec a}(A)$ and $S_{\vec b}(B)$ of the spins of
the two particles along arbitrary unit vectors $\vec a$ and $\vec
b$ \cite{Bell}. The old Quantum Mechanics prediction, based on the
spin state of Eq.˜ (\ref{entanglement}), was $ \langle S_{\vec
a}(A)S_{\vec b}(B)\rangle =-\frac{\hbar^2}{4}\vec a\cdot\vec b$,
which is the maximal correlation (in absolute value) that can be
achieved. In fact, the original EPR paradox requires strictly
maximal correlations, since it is about completeness and
determinism, being formulated for the objective ``physical
reality" (as we have seen, the measurement on A should imply a
{\it certain} prediction of the measurement on B). These are
theoretical problems, impossible to test directly due to the
unavoidable experimental errors. On the other hand, Bell's version
of the EPR experiment was explicitly aimed at testing the hidden
variables solution of the paradox \cite{Bell}, and not the EPR
paradox itself.

In fact, we have seen that the SM is EPR-paradox-free due to the
photons uncertainty. This is a theoretical result. It is not
difficult to see that the SM also predicts EPR correlations close
to those calculated by the old quantum mechanics approach that
agreed with the experimental data. In fact, allowing for the
radiation of additional photons, the correlations will be smaller
than maximal. However, we have already seen that the correction
will imply some suppression factors at least proportional to some
power of the fine structure constant. Moreover, the selection of
the events where A and B appear in opposite directions implies a
further, severe reduction of the phase space available for the
additional photons, whose transverse momenta should add up to zero
within the small solid-angle uncertainty given by the cross
section of the detector divided by the distance from the
production point. For this reason, the contributions to the
correlations from the diagrams involving additional photons  are
expected to be small compared with the experimental errors, and
will not spoil the previous agreement with the data. Such
correlations, although they are not strictly maximal, may still be
used in Quantum Computing and Quantum Information Theory
\cite{Laloe}.

The EPR correlations are usually interpreted as a sign of some
non-locality in themselves, since they cannot be justified by a
deterministic local theory based on additional ``hidden variables"
\cite{Bell,Laloe}. Such a supposed non-locality would now be less
problematic, since we know that locality is respected by all the
elements of physical reality; according to the EPR criterion, this
is sufficient to save the consistency of the theory (which is
challenged by the measurement problem anyway \cite{Laloe}).
Nevertheless, I think that the views that I have presented here
could suggest the need for a QFT approach to the whole problem of
locality. In fact, the original motivation for introducing hidden
variables in the study of the EPR correlations was the EPR
argument. Now this motivation has disappeared. Without introducing
hidden variables, should QFT be considered non-local invoking
``Bell's theorem" \cite{Bell,Laloe}? This question will possibly
require an approach to the measurement problem and in general to
the interpretation of QFT, and will be discussed elsewhere.

{\bf Conclusions.} The QFT description of Particle Physics has
been shown to be protected from the original EPR paradox by the
local gauge symmetry. This corresponds to the fact that it allows
for the creation of an undetermined number of photons in any
interaction, in particular in coincidence with the observed
particles in an EPR experiment. This result is particularly
important since it does not depend on the interpretation of the
theory, and it removes one of the most disturbing paradoxes of the
quantum theory, the violation of ``local realism". On the other
hand, the QFT description, which is not limited to the
``entangled" wave function with a definite number of particles,
can fulfill the EPR criterion of completeness (although it can
hardly be considered to be the ultimate ``theory of everything",
e.g.˜ it does not describe gravity). This solution would be
confirmed by the observation of an apparent symmetry violation in
a single event in an EPR experiment. On the other hand, the EPR
correlations are expected to be smaller than those calculated by
ignoring the soft photons, but in the case of the actual
experiments inspired by the work of Bell the correction is
expected to be small, so that the agreement of the Quantum Theory
with the present data is not spoiled. Such EPR correlations are
usually thought to be in themselves a sign of some ``quantum
nonlocality". This residual problem possibly depends on the
interpretation of the quantum theory and deserves further
research, that could profit from the views that I have presented
here.

It is a pleasure to thank Humberto Michinel for stimulating
discussions and help, and Esther P\'erez, Rafael A. Porto, Uwe
Trittmann and Ruth Garc\'\i a Fern\'andez and Rebecca Ramanathan
for useful comments.

\bibliography{pureprp}

\end{document}